\documentclass[aps, prd, twocolumn, superscriptaddress, nofootinbib]{revtex4}
\usepackage{graphicx}
\usepackage{dcolumn}
\usepackage{amssymb}

\begin{document}

\newcommand{\Eq}[1]{\mbox{Eq. (\ref{eqn:#1})}}
\newcommand{\Fig}[1]{\mbox{Fig. \ref{fig:#1}}}
\newcommand{\Sec}[1]{\mbox{Sec. \ref{sec:#1}}}

\newcommand{\PHI}{\phi}
\newcommand{\PhiN}{\Phi^{\mathrm{N}}}
\newcommand{\vect}[1]{\mathbf{#1}}
\newcommand{\Del}{\nabla}
\newcommand{\unit}[1]{\;\mathrm{#1}}
\newcommand{\x}{\vect{x}}
\newcommand{\ScS}{\scriptstyle}
\newcommand{\ScScS}{\scriptscriptstyle}
\newcommand{\xplus}[1]{\vect{x}\!\ScScS{+}\!\ScS\vect{#1}}
\newcommand{\xminus}[1]{\vect{x}\!\ScScS{-}\!\ScS\vect{#1}}
\newcommand{\diff}{\mathrm{d}}

\newcommand{\be}{\begin{equation}}
\newcommand{\ee}{\end{equation}}
\newcommand{\bea}{\begin{eqnarray}}
\newcommand{\eea}{\end{eqnarray}}
\newcommand{\vu}{{\mathbf u}}
\newcommand{\ve}{{\mathbf e}}

        \newcommand{\vU}{{\mathbf U}}
        \newcommand{\vN}{{\mathbf N}}
        \newcommand{\vB}{{\mathbf B}}
        \newcommand{\vF}{{\mathbf F}}
        \newcommand{\vD}{{\mathbf D}}
        \newcommand{\vg}{{\mathbf g}}
        \newcommand{\va}{{\mathbf a}}


\title{Time delays across saddles as a test of modified gravity}

\newcommand{\addressImperial}{Theoretical Physics, Blackett Laboratory, Imperial College, London, SW7 2BZ, United Kingdom}

\author{Jo\~{a}o Magueijo}
\affiliation{\addressImperial}
\author{Ali Mozaffari}
\affiliation{\addressImperial}

\date{\today}

\begin{abstract}
Modified gravity theories can produce strong signals
in the vicinity of the saddles of the total
gravitational potential. In a sub-class of these models 
this translates into diverging time-delays for echoes crossing
the saddles. Such models arise from the possibility that gravity might 
be infrared divergent or confined, and if suitably designed 
they are very difficult to rule out. We show that Lunar Laser Ranging 
during an eclipse could probe the time-delay effect within meters of 
the saddle, thereby proving or excluding these models. Very Large 
Baseline Interferometry, instead, could target 
delays across the Jupiter-Sun saddle. Such experiments would shed
light on the infrared behaviour of gravity and examine the puzzling 
possibility that there might be well-hidden regions of strong gravity and even 
singularities inside the solar system. 
\end{abstract}


\maketitle


Even though it is inevitable that Einstein's theory of general 
relativity (GR) will not be the final word, it is telling that
almost a century after its proposal
all theories trying to supersede it  have been ruled 
out or remain beyond detection~\cite{will,willcent,Clifton1}.
Nonetheless, it is precisely the  experimental
misfortunes of ``modified gravity'' that prove the 
strength of GR, 
so it is important to keep pushing the boundaries, constructing and
observationally disproving
new possibilities. A further motivation derives 
from attempts to combine quantum theory and general relativity, a logical
(if not an empirical) necessity. 
Such efforts invariably lead to corrections 
to GR, often at energy scales beyond the reach of 
current experiment, but not always. Finally, 
we should never forget that face value---taking into account only the
matter sources that we do see---the observational status of GR 
in astrophysics and cosmology is calamitous. This is 
usually blamed on our imperfect knowledge 
of the matter content of the Universe, and dismissed by introducing
new forms of invisible matter. But it
could well be that the discrepancies signal a breakdown in our 
understanding of gravity. 

Among the many theories attempting to extend GR some have tried to 
address the last issue, doing away with the need for non-visible or 
``dark'' matter to explain anomalies at galactic, cluster
and cosmological levels (see e.g.~\cite{Milgrom:1983ca,Famgaughrev}). 
Such theories have been labelled 
``MOdified Newtonian Dynamics'' (MOND), even though
they have now been embedded into fully relativistic field 
theories (e.g.~\cite{teves,BSTV,aether,aether1,Bruneton,Milgrom:2009gv,Milgrom:2010cd}).
In all of them new effects are triggered below an acceleration scale, 
$a_0\sim 10^{-10}\unit{ms}^{-2}$, a property suggested by the phenomenology.
The fact that MONDian behaviour
is physically triggered by an acceleration scale 
does not preclude writing $a_0$ in terms of length scale:
\be\label{l0}
L_0=\frac{c^2}{a_0}
\ee
for which $a_0$ could be a proxy. It is interesting that 
$L_0\sim 30,000\unit{Mpc}$ is of the order of the current horizon/Hubble 
radius. Nonetheless, it is the onset of low acceleration that triggers
new effects. MONDian theories do not have preferred frames 
and do not break diffeomorphism 
invariance; yet new effects emerge in the non-relativistic approximation
when the {\it total} Newtonian force per unit mass 
falls below $a_0$. 
For this reason one may expect the presence of ``MONDian habitats'' 
in the small regions encasing the saddle points of the gravitational
potential in the Solar System, the points where the Newtonian force 
vanishes~\cite{bekmag}. The prospect of
a MONDian saddle test has motivated extensive 
work~\cite{ali,bevis,companion,aliqmond,TypeII,aliscale}, with Lisa Pathfinder
(LPF) in mind, but not only~\cite{Grail}.  

One of the most powerful tests of GR, by now elevated to
the category of ``classical'' test, employs
the echo time-delay effect. By flashing ``light'' (usually a radio wave)
at a distant object, and catching its reflected ``echo'', 
one measures a distinctive delay, if its path intersects
a strong gravitational field. This so-called Shapiro effect was first 
detected with the radar echo off Venus in superior conjunction, leading to
stringent constraints on the $\gamma$ PPN parameter~\cite{will}. 
Since then the observational front has improved very fast.
These tests
use the fact that for a large class of theories (metric theories)
the travel time is given by
\be\label{delay}
t=\frac{1}{c} \int {\left(1-2\frac{\Phi}{c^2}\right)} dz\; ,
\ee
where $\Phi$ is the total gravitational potential, as obtained in the 
non-relativistic limit. Whilst the delay along a single path may be gauged
away, the {\it variation} in delays along neighbouring paths is operationally
meaningful, and constitutes a bona fide observational target.

\begin{figure}
\begin{center}
\scalebox{0.5}{\includegraphics{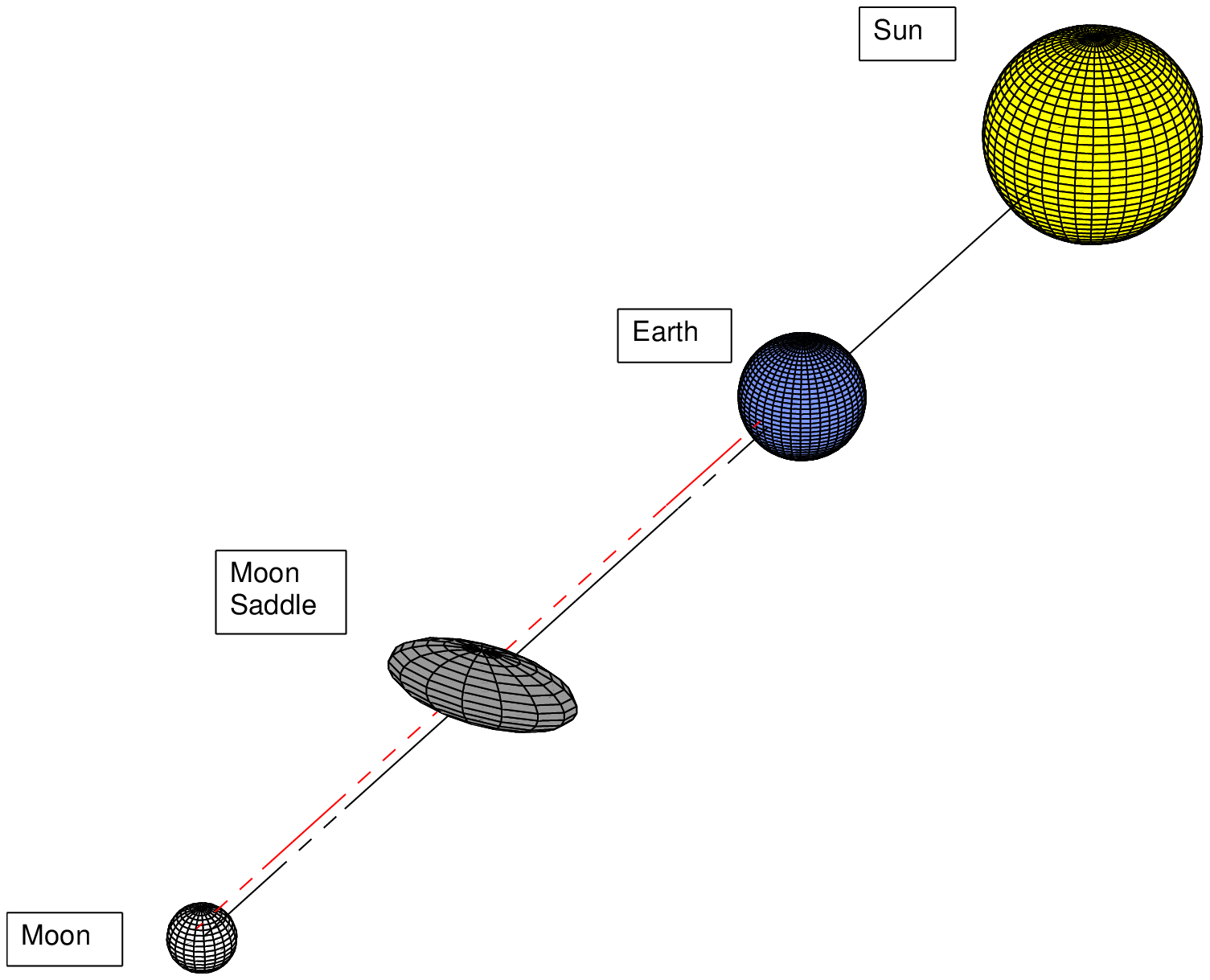}}
\scalebox{0.5}{\includegraphics{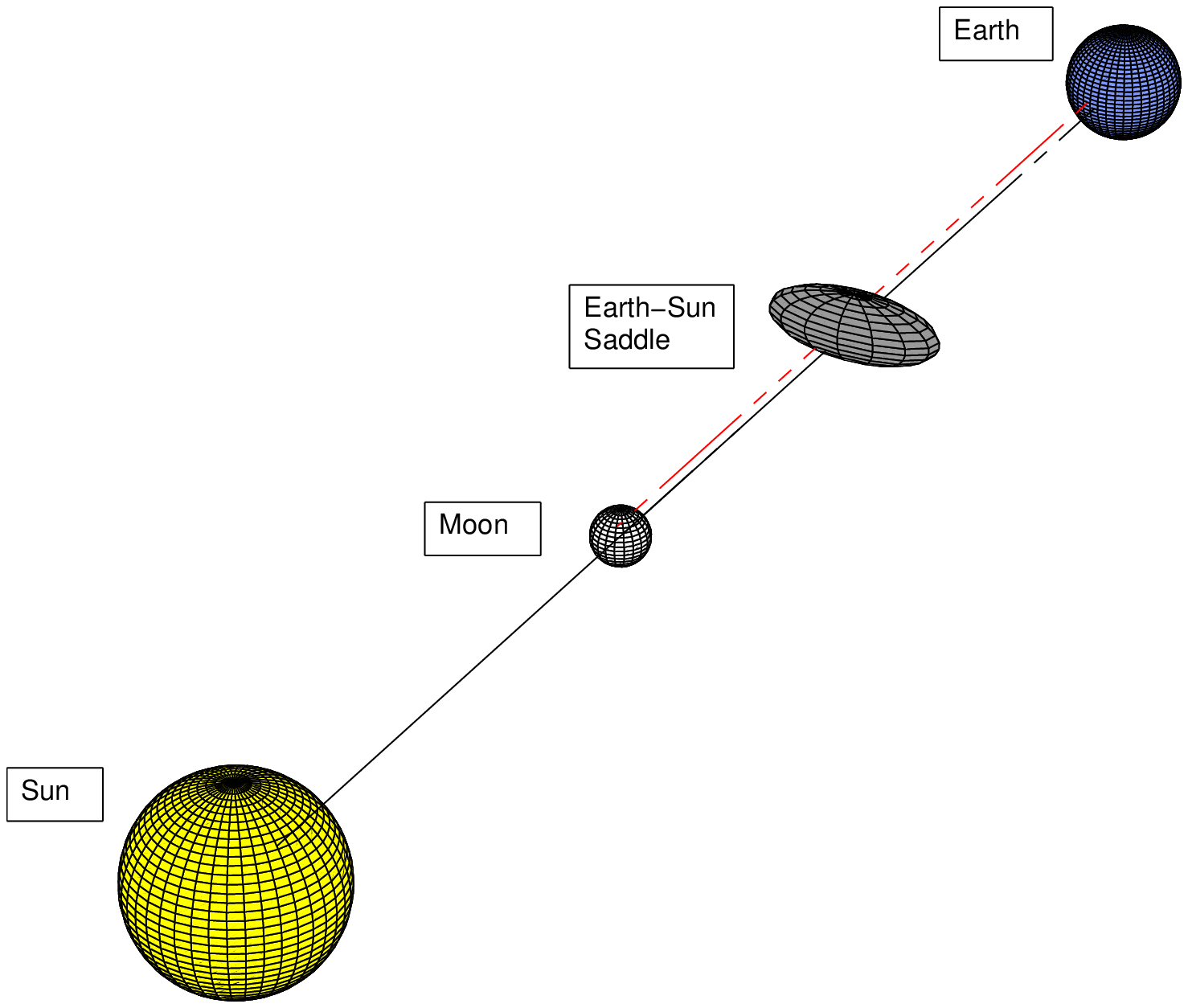}}
\caption{\label{scheme}{Schematic depiction of the geometry of an LLR 
saddle test during a lunar eclipse (top) and a solar (bottom) eclipse, 
with the lunar
(t) and the Earth-Sun (b) saddle regions represented, and the line of sight shown 
in red.}}
\end{center}
\end{figure}

Lunar Laser Ranging (LLR) is a major asset, among other fields, 
in gravitational physics (see for 
example~\cite{llr,murphy} as well as~\cite{Braxmaier:2011ai,Selig:2012zf}
where the effect on MONDian theories in also considered). Using
lunar retroreflectors one may time very accurately echoes of
sharp laser signals. Progress has been made steadily and precisions 
of tens of picoseconds, corresponding to distances of around 1~mm, 
can now be achieved~\cite{murphy}. 
It is immediately obvious (see Fig.~\ref{scheme}) that in principle 
LLR could probe the 
Moon saddle during a Lunar eclipse, sensing a possible MONDian 
time-delay effect.  Given that the Sun-Earth saddle
is within the Lunar orbit, it turns out that it can also be targeted 
by LLR during a Solar eclipse (c.f. Fig.~\ref{scheme}). 
In practice a LLR saddle test requires the
correct vantage point on Earth or even in its orbit. Although it's obvious
that the correct alignment is possible during an eclipse 
{\it if other perturbations are ignored}, the matter is less obvious
in the more realistic solar system. However, it turns out that if 
perturbations to the 3-body set up are taken into account, the exact 
location of the vantage point does changes, but not its presence on Earth or 
its orbit. Jupiter shifts the saddle location by 
at most 8~km, Saturn by 0.5~km. The galaxy displaces the saddle by about 
a meter and the extra-galactic field by a meter at most 
(see discussion in Section VII of~\cite{bekmag}). In contrast the moon can
shit the Earth-Sun saddle by as much as 6,000~km (see Fig. 8 in~\cite{bevis}).

Very Large Baseline Interferometry (VLBI) is another asset in
gravitational science (see e.g.~\cite{vlbi1}), providing 
another strategy for probing delays across saddles. It correlates
images of the same object (say, a quasar) as obtained in different 
continents. A set up could be arranged in which one light ray
goes through a saddle whilst the other  
(a few thousand kilometers away) does not. VLBI has 
the advantages that it could be used to probe other saddles,  
(e.g. the rather large Sun-Jupiter saddle), and that it relies only
on the presence of a source behind the saddle, rather than on the
vagaries of eclipses. 
In spite of potential difficulties, a saddle test with LLR or VLBI
could be carried out without the logistic overheads associated with the 
LPF saddle extended mission. 

It turns out that the time-delay effect is negligible for the 
MONDian models usually taken as targets for LPF. But there are also models well
beyond the reach of LPF which predict a strong time-delay signal.
Therefore the experimental test examined in this paper 
is complementary to a LPF test. This is hardly surprising. 
LPF accelerometers are sensitive to tidal stresses, i.e. the {\it second 
derivatives} of the gravitational potential 
(they feel their variation on a given frequency range). 
A time-delay test, instead, 
is sensitive to the {\it integral} of the potential along the line
of sight (or rather, 
to its variation across neighbouring paths). It is therefore
natural to find complementarity between the two measurements. An
extension to LPF would
constrain modified gravity theories which do away with the need 
for dark matter. A time-delay measurement would instead probe theories
which encode the property that gravity is subject to confinement,
as we shall now see.

It has long been speculated that gravity might resemble QCD. 
This has been found in field theory, string theory, in 
studies invoking holographic correspondences and in the study of the 
renormalization group flow (for a sample of references scattered throughout
several decades see~\cite{Salam,Lipatov,Giddings,Bern,Evans}.
In particular studies of the renormalization group flow have shown a 
strong paralel between the two theories, suggesting that 
gravity could be asymptotically free
(or ``safe'') and reciprocally be subject to confinement, with a 
divergent strength at large distance, or low energy
(e.g.~\cite{safe,safe2,safe1}). The
latter is actually what happens in the presence of a negative 
cosmological constant, inducing an attractive force satisfying 
Hook's law (diverging like $r$). But more generally the issue
has been raised in the context of the renormalization group flow of
quantum general relativity, where it has been conjectured that
the degenerate fixed point, when lifted, contains divergent 
IR behaviour.  An equivalent implementation of this phenomenology 
is possible in the context of
MONDian theories, should these be released from their obligations as dark 
matter alternatives, but with a key property kept and exacerbated.
Existing MONDian theories already have the peculiarity that they 
enhance the strength of gravity in situations where the standard 
Newtonian force would become weak. Specifically, 
for astrophysical applications, as $F_N\ll a_0$, we have that the MONDian
force goes like $F_\phi\sim\sqrt{F_N}$, 
at least in spherically symmetric situations. Thus, the MONDian 
force still drops to zero with $F_N$, albeit slower. 
But what if it diverged instead?
For example, we could imagine  the ``dual'' behaviour
$F_\phi\propto 1/F_N$,
and once this  possibility is considered we could consider sharper 
divergences, such as exponentials,  or 
\be\label{dual}
F_\phi\sim\frac{1}{F^p_N}
\ee
with $p>0$ very large. If one is to avoid appealing to dark matter
clearly $a_0$ would have to be smaller than the usual one, but not
otherwise. It turns out that it would be very difficult to rule out 
theories of this sort, except for their echo and VLBI saddle delays, 
as now show.

We first briefly lay down the formalism for defining MONDian theories, 
without wedding ourselves to a specific formulation. 
As explained in~\cite{ali}, in spite of the  large number of 
MONDian theories, the expression for the non-relativistic
potential invariably satisfies 3 types of equations only
(which may be formally reduced to 2). 
The dynamics may be written as resulting from the  
usual Newtonian potential $\Phi_N$ and a ``fifth force'' 
field, $\phi$, responsible for MONDian effects, with  total potential 
$\Phi=\Phi_N+\phi$.  For ``Type I'' theories  
$\phi$ is ruled by a non-linear Poisson equation: 
\be \label{eqI}\nabla \cdot
\left(\mu(z)\nabla \phi\right) = \kappa G \rho, \ee where, for
convenience, we pick the argument of the free function function
$\mu$ as: \be
z=\frac{\kappa}{4\pi}\frac{\vert\nabla\phi\vert}{a_0} \ee where
$\kappa$ is a dimensionless constant.  
For ``Type II'' theories we have instead:
\be \label{eqII}\nabla^2 \phi =
\frac{\kappa}{4\pi}\nabla\cdot\left(\nu(v)\nabla \Phi_N \right)
\ee 
where the argument of free function $\nu$ is given by \be \label{argII}
v=
\left(\frac{\kappa}{4\pi}\right)^2\frac{\vert\nabla\Phi_N
\vert}{a_0} \; . \ee 
Should these theories serve their duties as dark matter alternatives
we should require
$\mu\sim z$ for $z\ll 1$ and  $\nu\sim 1/\sqrt{v}$ for $v\ll 1$,
and $a_0$ should be the usual MONDian acceleration. 
However these theories have an interest in their own right:
they may generally be regarded as theories with a preferred acceleration scale. 
More general functions $\mu$ or $\nu$ and values of $a_0$ should then
be considered. If we want to keep the alternative to dark matter rationale,
then the $a_0$ used here must be at least one order of magnitude smaller
than that employed in traditional MONDian theories. However, if we detach
these theories completely from that role, and if dark matter does exist and 
play a role in the dynamics, this is not true.

We illustrate our calculations using type II theories, because they're
simpler.
For the purpose of investigating confined gravity we shall consider
free functions which for $v\ll 1$ are power laws: 
\be\label{nun}
\nu\propto \frac{1}{v^n}\; .
\ee 
From (\ref{eqII}) and (\ref{dual}) we have
$p=n-1$ and so for $n>1$ (to be contrasted with
the MONDian $n=1/2$) we obtain confinement behaviour. For $n=3/2$ the theory 
mimics a negative cosmological constant 
in spherically symmetric situations. The dual behaviour suggested
in (\ref{dual}) with $p=1$ follows from $n=2$. Note that a negative Lambda
is the perfect dual to standard MOND behaviour. 
The potential $\phi$ diverges around a saddle if $n\ge 2$. However,
this only translates into a divergent time-delay at the saddle 
if $n\ge 3$, with $n=3$ representing
a  logarithmic divergence 

Even though parameter $\kappa$ will not appear in the final answer
for the time-delay, it is important for justifying an approximation
and setting a scale. The rationale for its appearance in (\ref{argII})
and proportionality constant in (\ref{eqII}) is as follows~\cite{ali}.
If $\nu\rightarrow 1$ at large argument then Newton's constant
$G$ is renormalized by $\kappa/(4\pi)$, and this should be small. 
But so that $F_\phi\sim a_0$ when $F_N\sim a_0$,
we should use (\ref{argII}) for argument of $\nu$, if  
$\nu\sim 1/\sqrt{v}$ is to be triggered at $v\sim 1$ in 
the usual MONDian theory. 
With (\ref{nun}) the same requirement becomes:
\be v=
\left(\frac{\kappa}{4\pi}\right)^{\frac{1}{n}}\frac{\vert\nabla\Phi_N
\vert}{a_0} \; , \ee 
with $\nu\propto 1/v^n$ triggered at $v\sim 1$. The region where 
field $\phi$ goes strongly MONDian is now
of size:
\be 
r_0=\frac{a_0}{A}{\left(\frac{\kappa}{4\pi}\right)}^{\frac{1}{n}}\; ,
\ee
where $A$ is the diagonal tidal stress at the saddle along the line connecting
the two bodies. 
However, the field $\phi$ is subdominant with respect to the Newtonian potential
until we get to a distance:
\be\label{rtilde}
{\tilde r}=\frac{a_0}{A}
\ee
from the saddle.
It is only inside this inner bubble that $\phi$ is both MONDian {\it and}
dominant.

For orientation purposes we first evaluate the delay effect in 
Newtonian theory. Introducing cylindrical coordinates,
$\{z,\rho,\theta\}$, the delay is obtained by integrating 
Eq.~(\ref{delay}) along paths of constant $\rho=b$. (This is valid
for eclipse LLR only; the geometry is more complex for VLBI.) 
In the linear approximation the Newtonian potential is
$\Phi^N=A{\left( -\frac{z^2}{2}+\frac{\rho^2}{4}\right)}$ (see for example
the discussion in~\cite{bekmag}, for the validity of this approximation). 
For the Earth-Sun
saddle the tidal stress is
$A\approx 4.6\times 10^{-11}\rm{s}^{-2}$. Evaluating 
(\ref{delay})  over $z\in (-L/2,L/2)$ in these coordinates and with 
this potential is straightforward algebra, and the 
only term that varies with impact parameter $b$ (and so is observable)
is $\Delta t =- \frac{L}{c^3}\frac{Ab^2}{2} $. Even if the linear
approximation were valid throughout the whole flight (an assumption
which provides an upper bound on the real effect), we'd get
variations of the order of $10^{-16}$s for impact parameters 
$b\sim 1000$~km (with a delay at the center with respect to outer
trajectories). For the Jupiter saddle the effect is even smaller
since $A\sim 1.8\times 10^{-14}\rm{s}^{-2}$. The Newtonian delay is 
therefore negligible.

We now repeat this calculation for type II theories, mimicking the 
calculation of $\phi$ in~\cite{aliqmond}, for a free function of
form (\ref{nun}). The potential satisfies ansatz:
\be\label{phi}
\phi=-\frac{a_0^n}{A^{n-1}}\frac{1}{r^{n-2}}(f_0+f_2\cos(2\psi)+f_4
\cos(4\psi)+...)
\ee
where the parameters $f$ have to be determined numerically. 
A particularly simple case 
follows from $n=4$.
The integration should be performed along $z$ within the region of strong
MONDian behaviour, delimited by $r_0$ as defined above.
For $n=4$ the integration can be carried out explicitly, with the 
3 terms in~(\ref{phi}) integrating into:
\be
\Delta t\approx \frac{4a_0^4}{c^3A^3 b}\left(f_0\arctan{x}
-\frac{f_2x}{1+x^2} +f_4\frac{x(1-x^2)}{(1+x^2)^2}\right)\; ,
\ee
(where $x=r_0/b$), where the first term always dominates.
If we can assume that $r_0\gg b$ (always true if $\kappa\ll 1$), 
this becomes asymptotically:
\be\label{dt}
\Delta t\approx \frac{2\pi f_0}{b}\frac{a_0^4}{c^3A^3}\; ,
\ee
(i.e. only the first term survives).
Parameter $\kappa$ does not appear in the final answer, as long
as it is small enough to justify taking $r_0/b\rightarrow \infty$. 
For more general $n$ the calculation is more elaborate, 
leading to the asymptotic result ($b/r_0\rightarrow 0$):
\be\label{deltat}
\Delta t\approx \frac{C}{b^{n-3}}\frac{a_0^n}{c^3A^{n-1}}\; ,
\ee
where $C$ is given by:
$$
C=\frac{2\sqrt{\pi}\Gamma\left(\frac{n-3}{2}\right)}
{\Gamma\left(\frac{n-2}{2}\right)}
\left(f_0+\frac{n-4}{n-2}f_2+\frac{(n-4)(n-6)}{n(n-2)}f_4\right)\; .
$$
As we see, $n=4$ is a particularly simple limit of this expression.
We note that for $n>3$ the relative time variation at the saddle
diverges, decreasing as a power-law in $b$ as we move away from 
the ``bull's eye''.  In principle the constant $C$ can be positive 
or negative, leading to a delay or an advance at the bull's eye,
but we shall call it delay for definiteness.

\begin{figure}
\begin{center}
\scalebox{0.7}{\includegraphics{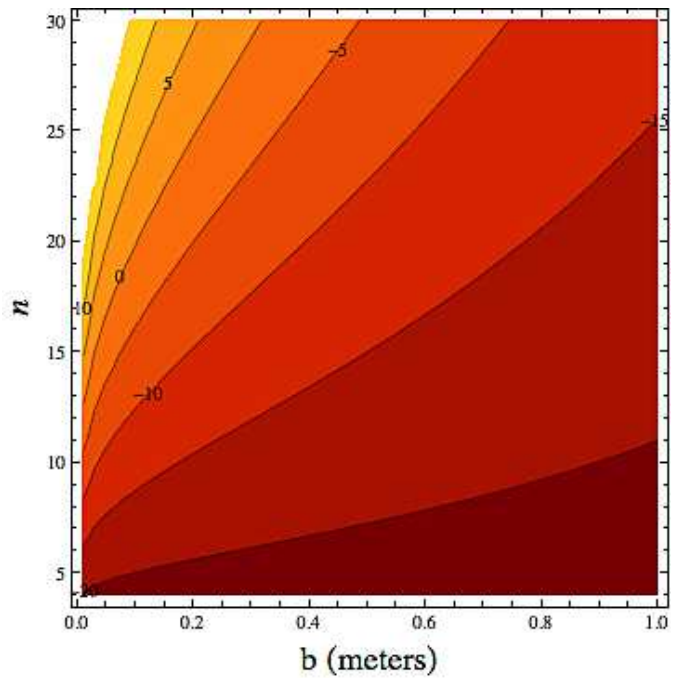}}
\caption{\label{ESfig}{The Log10 of the delay in picoseconds, as a function
of impact parameter $b$ and exponent $n$ for the Earth-Sun saddle, as probed,
say, by LLR during a Solar eclipse. As we can see the delay goes very quickly
from very small to very large. Realistically, with current technology,
only an integrated effect might be observable, and even then only for 
large $n$.}}
\end{center}
\end{figure}

In spite of this ``divergence''the observational implications are less 
dramatic than might be expected. 
Using the cosmological length scale $L_0$ defined in Eqn.(\ref{l0}) and 
the strong MOND bubble scale $\tilde r$ (in~Eqn.(\ref{rtilde}))
we can rearrange Eq.(\ref{dt}) in the suggestive form:
\be\label{dt1}
\Delta t\approx C \, \frac{b}{c}\frac{\tilde r}{L_0}
{\left(\frac{\tilde r}{b}\right)}^{n-2}\; .
\ee
The first factor (besides $C$) is just the time it takes to cross
the region closest to the saddle.
Unlike time-delays caused by the Sun this is small, because the 
distances involved are small: $b$ should be smaller than $\tilde r$ and 
${\tilde r}\sim 2.2\unit{m}$ for the Earth-Sun saddle and
${\tilde r}\sim 5.5\unit{km}$ for the Jupiter saddle. In addition the 
second factor relates the MOND bubble size to the horizon scale, introducing 
a tiny factor. Therefore, even though the third factor predicts a divergence, 
this will happen very close to the saddle and be observable only for very 
steeply diverging functions.

\begin{figure}
\begin{center}
\scalebox{0.7}{\includegraphics{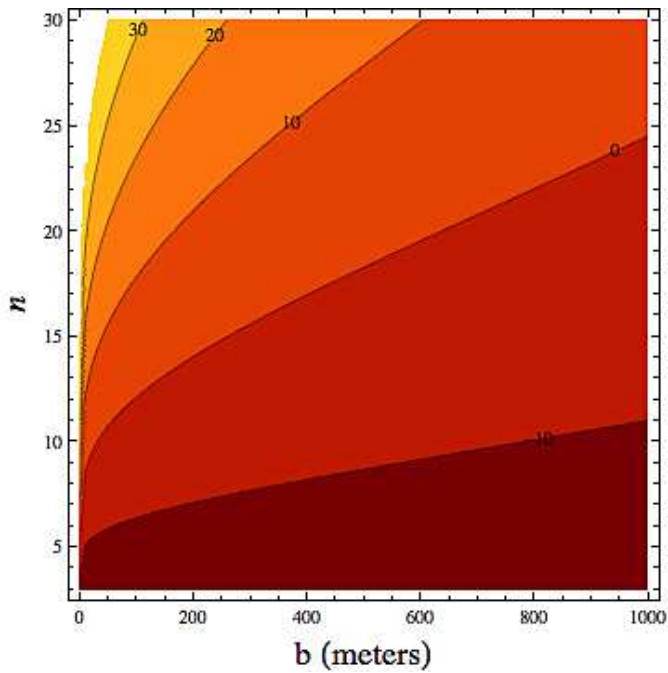}}
\caption{\label{JSfig}{The Log10 of the delay in picoseconds, as a function
of impact parameter $b$ and exponent $n$ for the Jupiter-Sun saddle, 
as potentially probed by VLBI. Again the delay goes very quickly
from very small to very large. Here the boundary is closer to realistic
experimental parameters, and lower exponents $n$ come within reach.}}
\end{center}
\end{figure}

We spell out this expectation in Figs.~\ref{ESfig} and~\ref{JSfig},
which describes the situation for the Earth-Sun saddle 
(as a target for LLR during a Solar eclipse) and 
the Jupiter-Sun saddle (as a target for VLBI), respectively. 
In both of these figures we have plotted the (base 10) logarithm of the
time-delay in picoseconds, as a function of the impact parameter $b$
(in meters) and the exponent $n$ used in free-function $\nu$. 
In both cases we observe a very abrupt
transition from the very small to the very large, with the 
contour labeled zero denoting the rough borderline for observability
with current technology.
Typically the Earth saddle would have to be probed closer to a meter 
and even then 
assuming large values of $n$ (in the range 20-30). The Jupiter saddle
might be more forgiving, and small values of $n\sim 5$ could come within
reach for $b$ of the order of a meter, with $n\sim 20$ still constrained
even for impacts of the order of a kilometer. In all fairness we cannot
be overenthusiastic about the detectability of this effect in the first 
setting, where with current technology it would be seen at best as an 
integrated effect (the wavepackets often have a width of about 200 meters).
The next generation of lunar retroreflectors could be necessary.
The second situation might be more hopeful. We illustrated our 
conclusions with the Earth-Sun saddle during a Solar eclipse
but similar results apply for the Moon saddle, as targeted by LLR during
a Lunar eclipse. Likewise what we have shown for the Jupiter saddle
has a closely related counterpart with the Saturn saddle. 
Incidentally, Eq.~(\ref{dt1}) can be used to prove that the effect 
for standard MONDian functions ($n=1/2$) is negligible ($\Delta t
\sim 10^{-34}\unit{s}$ with $b$ about a meter from the saddle). Likewise
it can be shown that the functions considered here would have
negligible effect for a LPF test.

What about other MONDian theories? As an example, we briefly discuss what
was labeled type I theories in~\cite{ali}. For these the 
situation is more complex due to the well known presence 
of a curl field, softening  the divergence~\cite{bekmag,ali}. 
Under strict spherical symmetry this field vanishes, so although this
is not applicable to a saddle, we can gain some intuition. 
Parametrizing the free-function in (\ref{eqI}) as: 
\be
\mu\propto \frac{1}{z^m}
\ee 
for $z\ll 1$, we see that {\it ignoring the curl field}, the exponent $p$
in (\ref{dual}) is  $p=1/(m-1)$, so that $m>1$ becomes the condition
for confining behaviour. Now, $m=3$ is equivalent to a negative 
cosmological constant $\Lambda$, and  $m=2$ leads to a perfect dual
($p=1$). However, conclusions about the conditions for a 
divergence at the saddle are more subtle, because the magnetic field cannot 
be neglected. If this were the case, then $1< m\le 2$ would lead to a diverging 
$\phi$, and $1< m < 3/2$ to a diverging delay. However, now
we can only try out the more flexible ansatz:
\be
\phi=-C_1\frac{1}{r^\alpha}(f_0+f_2\cos(2\psi)+f_4
\cos(4\psi)+...)
\ee
and search for a solution numerically (using techniques presented, in 
a different context, in~\cite{genmu}). We find for $m=1.1$, for example, 
that $\alpha=1.25$ (instead of $\alpha=9$, expected if the curl field
could be neglected). Thus, these theories are even more difficult to 
constrain than type II. The situation is similar with type III (which
also have a curl field).

It is interesting that something as dramatic as this divergence can
be so elusive. Furthermore, we have only solved the problem to linear
order, and many questions can be raised beyond the scope of the
calculation presented in this paper. 
For example, if gravity is confined and infrared divergent, as envisaged 
here, could there be a singularity at the saddle? If so, would 
this singularity be naked, or rather, would there be a horizon? 
Whilst a positive 
answer to the first question is plausible, the answer to the second 
question is far from obvious. In both cases the detectability of the 
time-delay effect for the free functions used above, as calculated
here, is unlikely to improve. The field is not attractive, 
so the usual arguments about accretion disks and X ray emission do 
not apply (even considering, e.g. the solar winds). One may think
it odd that naked singularities
or horizons could be floating around in the Solar system, but in practice the regions
where such extreme behaviour is felt are very small, and they could pass
unnoticed. 
 
Of course one could fluff up the divergence region by introducing functions 
of the form:
\be
\nu=\frac{1}{(v-1)^n} 
\ee
in type II theories, for example. 
Then, the non-linear regime would be entered close 
to the ellipsoid $z^2 +\rho^2/2={\tilde r}^2$, and
depending on the details of the full relativistic theory, this 
could signal the formation of a horizon or a naked singularity.
Either way,
assuming LLR geometry, any photons with 
$b<2{\tilde r}$ would be lost, i.e. they would have an infinite 
time-delay. 
Close to the disk defined by $b=2{\tilde r}$ the time-delay would 
diverge as:
\be
\Delta t = C' \frac{\tilde r}{c}\frac{\tilde r}{L_0}
\left(\frac{\tilde r}{b-2{\tilde r}}\right)^{n-2}
\ee
(written in a format to allow easy comparison with (\ref{dt1})).
We would now need to be glued to the ``horizon'' for the effect to be 
measurable, unless $n$ is very large. However, we would 
also have a ``black spot'', comprising the disk $b<2{\tilde r}$.
This rather extreme free-function is the only possibility we found for 
rendering these theories more tangible, and clearly a large $a_0$
is then promptly ruled out.

In summary, we hope that in this paper we have stressed the radical
difference between the gravitational physics probed by LLR or VLBI on 
the one hand, and LPF on the other, regarding saddle points.
With LPF one probes second derivatives of the potential, locally.
With LLR and VLBI one probes the integral of the potential, at the end
points, as a cumulative effect. Therefore with LPF, for standard MONDian
functions, we find a distinctly changing tidal stress at the saddle 
(to be contrasted 
with an essentially DC Newtonian background). We cannot realistically 
get close to
the saddle, but even far out we can expect signals with large SNRs. 
With a delay test
we can potentially probe the region very close to the saddle; however, 
the predicted effects for
standard MONDian functions are tiny. Nevertheless, 
we become sensitive to functions which diverge at low accelerations, 
associated with confinement and strong infrared behaviour for gravity.
Such theories predict  extreme behaviour very close to the 
saddle, raising the possibility of singularities. They are beyond 
the reach LPF and they do not purport
to present an alternative to dark matter. However they have an interest
in the own right, and are targets for a time-delay test as performed
by the current or next generation of lunar retroreflectors, and by VLBI.

{\bf Acknowledgments} We thank Marshall Eubanks and Jacob Bekenstein
for first bringing the potential of LLR to our attention, and 
A. Bonanno, D. Litim, T. Murphy and K. Stelle for comments. 
JM is funded by an STFC consolidated grant and 
AM by an STFC studentship.  Some of the numerical work was carried out on 
the COSMOS supercomputer, which is supported by STFC, HEFCE and SGI.

\bibliography{references}

\end{document}